\documentclass[11pt,epsfig]{article}

\usepackage{amssymb}
\usepackage{graphicx}
\usepackage{color}
\usepackage{subfigure}

\begin{document}

\baselineskip 6mm
\renewcommand{\thefootnote}{\fnsymbol{footnote}}


\newcommand{\nc}{\newcommand}
\newcommand{\rnc}{\renewcommand}



\newcommand{\tcb}{\textcolor{blue}}
\newcommand{\tcr}{\textcolor{red}}
\newcommand{\tcg}{\textcolor{green}}


\def\be{\begin{equation}}
\def\ee{\end{equation}}
\def\ba{\begin{array}}
\def\ea{\end{array}}
\def\bea{\begin{eqnarray}}
\def\eea{\end{eqnarray}}
\def\nn{\nonumber\\}


\def\ct{\cite}
\def\la{\label}
\def\eq#1{(\ref{#1})}


\def\a{\alpha}
\def\b{\beta}
\def\g{\gamma}
\def\G{\Gamma}
\def\d{\delta}
\def\D{\Delta}
\def\e{\epsilon}
\def\et{\eta}
\def\ph{\phi}
\def\Ph{\Phi}
\def\ps{\psi}
\def\Ps{\Psi}
\def\k{\kappa}
\def\l{\lambda}
\def\L{\Lambda}
\def\m{\mu}
\def\n{\nu}
\def\th{\theta}
\def\Th{\Theta}
\def\r{\rho}
\def\s{\sigma}
\def\S{\Sigma}
\def\ta{\tau}
\def\o{\omega}
\def\O{\Omega}
\def\pr{\prime}


\def\half{\frac{1}{2}}

\def\goto{\rightarrow}

\def\na{\nabla}
\def\grad{\nabla}
\def\curl{\nabla\times}
\def\div{\nabla\cdot}
\def\pa{\partial}
\def\fr{\frac}

\def\bra{\left\langle}
\def\ket{\right\rangle}
\def\lb{\left[}
\def\lc{\left\{}
\def\ls{\left(}
\def\lp{\left.}
\def\rp{\right.}
\def\rb{\right]}
\def\rc{\right\}}
\def\rs{\right)}

\def\vac#1{\mid #1 \rangle}


\def\td#1{\tilde{#1}}
\def\check{ \maltese {\bf Check!}}


\def\Tr{{\rm Tr}\,}
\def\det{{\rm det}}
\def\text#1{{\rm #1}}


\def\bc#1{\nnindent {\bf $\bullet$ #1} \\ }
\def\ch {$<Check!>$ }
\def\ss {\vspace{1.5cm}}
\def\inf{\infty}

\begin{titlepage}

\hfill\parbox{5cm} { }

\vspace{25mm}

\begin{center}
{\Large \bf On black hole thermodynamics  with a momentum relaxation}

\vskip 1. cm
  {Chanyong Park$^{a,b}$\footnote{e-mail : chanyong.park@apctp.org}}

\vskip 0.5cm

{\it $^a\,$ Asia Pacific Center for Theoretical Physics, Pohang, 790-784, Korea } \\
{\it $^b\,$ Department of Physics, Postech, Pohang, 790-784, Korea }\\

\end{center}

\thispagestyle{empty}

\vskip2cm


\centerline{\bf ABSTRACT} \vskip 4mm

\vspace{1cm}

We investigate black hole thermodynamics involving a scalar hair which is dual to a momentum relaxation of the dual field theory. This black hole geometry is able to be classified by two parameters. One is a momentum relaxation and the other is a mass density of another matter localized at the center. Even though all parameters are continuous, there exists a specific point where its thermodynamic interpretation is not continuously connected to the one defined in the other parameter regime. The similar feature also appears in a topological AdS black hole. In this work, we show why such an unusual thermodynamic feature happens and provide a unified way to understand such an exotic black hole thermodynamically in the entire parameter range.

\vspace{2cm}


\end{titlepage}

\renewcommand{\thefootnote}{\arabic{footnote}}
\setcounter{footnote}{0}


\section{Introduction}

For the last decades, there were many attempts to understand various black hole's properties like black hole  thermodynamics \cite{Bekenstein:1973ur,Bardeen:1973gs,Hawking:1974sw,Gibbons:1977mu}. It still remains as one of longstanding physics problems to account for the microscopic origin for the area law of black hole entropy and its thermodynamics. In order to figure out the area law, new concepts like the entanglement entropy and holography have been applied to black hole physics. Recently, it is further refined into the AdS/CFT correspondence and provides a new paradigm to explain non-perturbative physics of a non-gravitational quantum field theory (QFT) defined at the AdS boundary \cite{Maldacena:1997re,Gubser:1998bc,Witten:1998qj,Witten:1998zw}. Recently, a scalar hairy black hole with a momentum relaxation has been paid much attention in order to study various transport coefficients like a finite DC conductivity \cite{Hartnoll:2007ih}-\cite{Khimphun:2016ikw}. Unlike an ordinary black hole, its thermodynamic properties are not clearly understood because of the existence of an extra scalar hair. In this work, we will clarify its exotic thermodynamic features and investigate carefully why such unusual properties appear after finding its correct thermodynamic interpretation.

In general, a black hole solution appears as a vacuum solution with a singularity at the center. Like collapsing of a star, the gravitational attraction concentrates all mass of a star into a tiny volume. When its size becomes smaller than the black hole horizon, the star evolves into a black hole. In this procedure, the outside of the horizon is characterized by a vacuum solution of the Einstein equation because there is no matter. Finally, the black hole settles down to a specific form of the Kerr-Newman black hole \cite{Kerr:1963ud,Newman:1965tw,Newman:1965my}, whose properties are classified by only several macroscopic variables called primary hairs. According to the no-hair theorem, possible hairs are restricted to the black hole's mass, charge and angular momentum. Due to the similarity of black hole physics to thermodynamics, these hairs can be reinterpreted as thermodynamic variables and satisfy the thermodynamic law.

Although the no-hair theorem is proved in a four-dimensional asymptotically flat geometry \cite{Heusler M}, there were many examples which break the no-hair theorem in a curved space. Typical examples are various scalar hairy black holes. Following \cite{Coleman:1991ku}, such a scalar hair can be classified into a secondary hair. Unlike the primary hair, the secondary hair is generated with the help of basic fields associated with primary hairs. In many examples, a scalar hair just modifies the form of the thermodynamic quantity caused by the primary hair rather than generates its own thermodynamic variable \cite{Taylor:2008tg}-\cite{Khimphun:2016gsn}. This fact means that the secondary hair does not alter the thermodynamic law. In the AdS/CFT context, a scalar hairy black hole usually plays an important role in figuring out various qualitative features of a strongly interacting system like superconductivity, the existence of the Fermi surface, strange metallic behavior, etc \cite{Hartnoll:2008vx,Hartnoll:2008kx,Huijse:2011ef,Hartnoll:2009ns}. In addition, it would be important to understand exotic thermodynamics of a scalar hairy black hole for enlarging our knowledge about nontrivial black hole physics.

In the gravity, primary hairs of a black hole are identified with thermal quantities after assuming that a black hole is a thermal system satisfying all thermodynamic laws. According to the AdS/CFT correspondence, such a black hole thermodynamics can be reinterpreted as that of the dual QFT \cite{Park:2013ana,Park:2013dqa,Park:2014gja}. Intriguingly a scalar hairy black hole, which is dual to a strongly interacting QFT with a momentum relaxation, reveals several exotic features. The first is that, in spite of the fact that all parameters are continuous, its thermodynamic interpretation at a specific point is not smoothly connected to that defined in the different parameter region. The second is that the thermal energy of the scalar hairy black hole does not vanish even at zero temperature. Similar features appear in a topological AdS black hole which is an AdS black hole with a hyperbolic boundary. 

In order to figure out why such unusual properties occur in a scalar hairy black hole, we instigate physical quantities of its dual QFT by using the holographic renormalization \cite{Henningson:1998gx}-\cite{Papadimitriou:2011qb}. In the AdS/CFT context, it is one of the essential tools to extract information about a strongly interacting system even at finite temperature. Although the holographic renormalization of a planar AdS black hole directly yields the exact same thermodynamic quantities obtained from black hole's thermodynamics, it is not generally true for an AdS black hole with a nontrivial boundary topology. In general, the holographic renormalization includes information of the vacuum as well as the thermal state, so that it would be a useful tool to comprehend exotic features of a scalar hairy black hole \cite{Emparan:1999pm}. In this work, we redefine thermal energies like internal and free energies correctly by subtracting the vacuum energies from the holographic renormalization results. Then, the resulting thermal energies automatically vanish in the zero temperature limit. After this redefinition, we find that the vacuum energy must be reinterpreted as a thermal energy at a specific parametric point. This explains why the first exotic feature discussed above arises. The same method can also apply to a topological AdS black hole, which gives rise to the correct thermodynamic interpretation consistent with black hole thermodynamics.

The rest of this paper is organized as follows: In Sec. 2, after constructing a scalar hairy black hole we discuss its thermodynamics after assuming the first law of thermodynamics. In this process, we expose several unusual features manifestly. Before resolving these issues, we show that a topological AdS black hole also has similar exotic features and figure out why such behaviors occur in Sec. 3.  Similarly, in Sec. 4 we investigate correct thermodynamics of a scalar hairy black hole with accounting for the origin of exotic features. We finish this work with some concluding remarks in Sec. 5.


\section{Einstein-scalar gravity}

Recently, a black hole geometry with a momentum relaxation has been widely investigated to obtain a finite DC conductivity of the dual QFT \cite{Horowitz:2012ky}-\cite{Blake:2013owa}. Since the momentum relaxation leads to an additional parameter to the black hole geometry, its thermodynamic interpretation becomes obscure. Furthermore, the black hole involving a momentum relaxation shows several exotic behaviors. Figuring out such unusual features would be important to understand its own black hole thermodynamics more clearly as well as some aspects of the dual field theory from the holographic point of view. In this section, we will construct a simple black hole solution involving a momentum relaxation and then show what kinds of outlandish feature occur in its thermodynamic interpretation.
To do so, let us consider a simple five-dimensional Einstein-scalar gravity with a $SO(3)$ global symmetry \cite{Mateos:2011ix,Mateos:2011tv,Koga:2014hwa,Kim:2015dna,Kim:2015wba}
\be		\la{act:Einsteinscalar}
S = \fr{1}{16 \pi G} \int d^{5} x \sqrt{-g} \lb {\cal R} - 2 \L - \half \d_{ab} \
\pa_\m \ph^{a} \pa^\m \ph^{b} \rb ,
\ee
where $\L = - 6/R^2$ indicates a negative cosmological constant. Due to the $SO(3)$ global symmetry denoted by indices, $a$ and $b$, the solutions depends only on the magnitude of the scalar field, $\ph = \sqrt{ \d_{ab} \ph^a \ph^b}$. Assuming that this system allows a static asymptotic AdS geometry with a planar boundary denoted by $\lc t , x^i \rc$ with $i=1,2,3$, then $\ph$ is generally given by a function of $x^i$ and $r$ where $r$ means a radial coordinate in the asymptotic AdS space. 

In order to find such an asymptotic AdS geometry, let us first concentrate on the scalar field equation near the boundary. On the AdS geometry, the scalar field is governed by 
\be
0 = \fr{1}{\sqrt{-g}} \pa_M \sqrt{-g} \pa^M \ph^a (r, x^i) .
\ee
Assuming that $\ph^a$ is independent of $x^i$, its asymptotic solution reads \cite{Gubser:1998bc,Witten:1998qj}
\be
\ph^a (r) = \ph_0^a \ \ls 1 + \cdots \fr{}{} \rs + \fr{ \bra {\cal O}^a \ket}{r^4} \ \ls 1 + \cdots \fr{}{} \rs ,
\ee
where $\ph_0^a$ and ${\cal O}^a$ appear as integral constants. On the dual field theory side, $\ph_0^a$ and ${\cal O}^a$ are identified with the source and vacuum expectation value of the dual scalar operator, ${\cal O}^a$, respectively. If $\ph^a$ also relies on $x^i$, its asymptotic expansion can be further extended into 
\be
\ph^a (r,x^i) = \ph_0^a (x^i) \ \ls 1 + \cdots \fr{}{} \rs + \fr{ \bra {\cal O }^a  (x^i) \ket  }{r^4}  \ \ls 1 + \cdots \fr{}{} \rs .
\ee
In general, it is not easy to find an analytic solution beyond the probe limit. However, in the simplest but nontrivial case with $\ph_0^a  (x^i) \sim x^i \d^a_i$ and $\bra {\cal O }^a  (x^i) \ket  =0$, the analytic solution has been known. Here we briefly summarize its geometric solution. Equations of motion derived from the above action are given by
\bea
&& {\cal R}_{\m\n} - \half g_{\m\n} {\cal R} + g_{\m\n}  \L
= T_{\m\n}  ,  \la{eq:Einstein} \\
&& 0 =  \fr{\d_{ab}}{\sqrt{-g}} \pa_\m  \ls \sqrt{-g} g^{\m\n} \pa_\n \ph^b \rs ,
\eea
where
\be
T_{\m\n} = \half \d_{ab} \pa_\m \ph^a \pa_\n \ph^b - \fr{1}{4} g_{\m\n} \d_{ab} \pa_\r \ph^a \pa^\r \ph^b .
\ee
Now, let us take the following linear scalar field profile with an appropriate normalization \cite{Mateos:2011ix,Mateos:2011tv}
\be		\la{an:linearscalar}
\ph^a = 2 \sqrt{m_{\ph}} \  x^i \d^a_i  ,
\ee
where $\sqrt{m_{\ph}}$ describes a momentum relaxation of the dual QFT and $m_{\ph} \ge 0$. Under the usual metric ansatz for a black hole
\be
ds^2 =    -  r^2 f(r)  dt^2 + r^2 d \vec{x}^2  + \fr{1}{r^2 f(r)} dr^2 ,
\ee
the above scalar profile automatically satisfies the equation of motion. Substituting these ansatz into the Einstein equation, it reduces to
\bea
0 &=&  r^3 f' + 4 r^2 f - 4 r^2 + 2 m_\ph  , \nn
0 &=&  r^4 f''+8 r^3 f'+12 r^2 f -12 r^2  +2 m_\ph ,
\eea
where the first and second equation describe a constraint and dynamical equation for an unknown metric function, $f$, respectively. The solution satisfying above two equations simultaneously is given by
\be			\la{res:generalblackhlsol}
f (r) =  1 - \fr{m_\ph}{r^2} - \fr{m}{r^4} ,
\ee
where $m_\ph$ and $m$ are two free parameters. This five-dimensional black hole can be easily generalized to other dimensions having an additional electric charge \cite{Kim:2015wba} . 

Before investigating thermodynamic properties of this black hole, it is worth noting that it allows us to take a negative value for $m$ because of the existence of a positive $m_\ph$. The range of $m$ should be restricted
to $m \ge - m_{\ph}^2 /4$. In the outside of this range, the resulting geometry is not a black hole but a space with a naked singularity at the center. When $m_\ph=0$, it becomes a usual Schwarzschild AdS (SAdS) black hole. In order to get a nontrivial $m$, we have to consider an additional matter localized in a very tiny region. If its size is smaller than the event horizon, $r_h=m^{1/4}$, the localized matter evolves into a black hole characterized by its mass, $m$. This implies that matter related to $m$ has nothing to do with the previous scalar field, $\ph^a$. 
In this case there is no matter outside the horizon, so the SAdS black hole describes a vacuum geometry. The black hole mass corresponds to the primary hair and satisfies the first law of thermodynamics. These pictures of an ordinary black hole have been well understood. From now on, we will investigate unusual aspects of the above black hole.

\subsection{Only with a momentum relaxation}

As mentioned before, $m$ and $m_{\ph}$ parameterize energies of two different matters. If there is no well localized matter ($m=0$), one cannot expect the existence of a black hole solution. This is true in a flat space. Without well localized matter a uniformly distributed scalar field, which is independent of the radial coordinate, cannot evolve to a black hole.  However, it is not the case in an AdS geometry. A uniformly distributed scalar field can make a black hole because of the nontrivial radial coordinate dependence of the background metric. Following \cite{Coleman:1991ku}, $m_\ph$ can be regarded as a secondary hair because the formation of the black hole crucially depends on the background geometry. For understanding more details, let us study how the curvature of the resulting space depends on the distribution of the scalar field.
From \eq{eq:Einstein}, the curvature scalar is given by 
\be
{\cal R} = \fr{10}{3} \L + \fr{6 m_\ph}{r^2} .
\ee
This result shows that the momentum relaxation causes a naked singularity at $r=0$ because the energy-momentum tensor of the scalar field diverges at the center. In order to obtain a regular geometry, this singularity should be hidden behind the black hole horizon which can be represented by the following black hole solution
\be
f (r) =  1 - \fr{m_\ph}{r^2}  ,
\ee
which is the exact same as \eq{res:generalblackhlsol} with $m=0$. The existence of the black hole horizon 
at $r_h=\sqrt{m_\ph}$ indicates that the scalar field uniformly distributed in the AdS space can evolve a scalar hairy black hole without well-localized matter.

Now, let us investigate thermodynamic properties of this black hole by using the thermodynamic law. The absence of a conical singularity at the horizon leads to the following Hawking temperature 
\be
T_H = \fr{r_h}{2 \pi}= \fr{\sqrt{m_\ph}}{2 \pi} , 
\ee
and the Bekenstein-Hawking entropy proportional to the area of the horizon reads
\be		\la{res:interenm0}
S_{BH} = \fr{ V}{4 G} r_h^3 = \fr{2 \pi^3  V}{G}  T_H^3  ,
\ee
where $V =  \int d^3 x$ denotes a regularized area of spatial directions.
Assuming that these black hole quantities satisfy the first law of thermodynamics, $d E = T_H d S_{BH}$,
its integration leads to the following internal energy 
\be                         \la{res:freeenm0}
E = \fr{3 V}{32 \pi G} r_h^4 = \fr{3 V}{32 \pi G} m_\ph^2 = \fr{3 \pi^3 V}{2  G} T_H^4 ,
\ee
which as expected satisfies the Stefan-Boltzman law. From these results, the free energy becomes
\be
F \equiv E - T_H S_{BH} =-  \fr{V}{32 \pi G} r_h^4  = -  \fr{\pi^3 V}{2 G} T_H^4.
\ee
If all these quantities satisfy the thermodynamics law, the derivative of the free energy with respect to the Hawking temperature should be reduced to the entropy with a minus sign
\be
S_{BH} = - \fr{d F}{d T_H} .
\ee
The above free energy really satisfies this thermodynamic relation. Another point we should notice is that this black hole solution has a well-defined zero temperature limit. In the zero temperature limit ($T_H \to 0$ and $m_{\ph} \to 0$), the internal energy corresponding to the thermal energy automatically vanishes. 

\subsection{With two kinds of matter}

Now, let us consider a more general case including two different matter. One is localized at the center and the other is uniformly distributed in the entire region. The black hole composed of these two types of matters in the AdS space is described by 
\be
f (r) =  1 - \fr{m_\ph}{r^2} - \fr{m}{r^4} .
\ee
Unlike the usual SAdS black hole, this black hole solution has two free parameters representing two different matters. In order to understand its thermodynamics, it is more convenient to represent $m$ in terms of $m_\ph$ and $r_h$
\be
m = r_h^2 \ls r_h^2  - m_\ph \rs .
\ee
The absence of the conical singularity determines the Hawking temperature to be
\be
T_H = \fr{r_h}{\pi} \ls 1 - \fr{m_\ph}{2 r_h^2} \rs .
\ee
When $m$ is positive, the horizon and Hawking temperature is restricted to $r_h > \sqrt{m_\ph}$ and $T_H > \sqrt{m_{\ph}}/2 \pi$. On the other hand, for $- m_{\ph}^2 /4 \le m < 0$ they are limited to $ \sqrt{m_{\ph}/2}\le r_h < \sqrt{m_\ph}$ and $0 \le T_H < \sqrt{m_{\ph}}/2 \pi$, respectively. In the latter case, the black hole has two horizons, inner and outer horizon, and $r_h$ indicates the outer horizon. Especially,  the extremal limit corresponding to the zero temperature limit appears at $r_h =\sqrt{m_{\ph}/2}$ with $m = - m_{\ph}^2 /4$. If $m < - m_{\ph}^2 /4$, there is no black hole horizon so that the resulting geometry shows a naked singularity at the center. 
Remember that the black hole composed of two matters allows a negative mass unlike a usual SAdS black hole. The similar situation also occurs in a topological black hole which is a SAdS black hole solution with a nontrivial hyperbolic boundary topology. In order to understand thermodynamics of the above black hole correctly, it is crucial to figure out thermodynamics of a topological black hole. We will revisit this issue in the next section.

In the rest of this section, we will discuss some unusual properties of the above black hole. The integration of the first law of thermodynamics yields  the following internal and free energy
\bea		\la{res:energieswithm}
E &=& \fr{3 V}{16 \pi G} r_h^2 \ls r_h^2 - m_\ph \rs = \fr{3 V}{16 \pi G} m , \nn
F &=& - \fr{V}{16 \pi G} r_h^2 \ls r_h^2 + m_\ph \rs .
\eea
At zero temeprature ($T_H=0$ and $r_h = \sqrt{m_\ph /2}$), the internal and free energy reduce to
\be			\la{res:shiftenergy}
E_{sh} = F_{sh}  = - \fr{3 m_\ph^2 V}{64 \pi G}  .
\ee
Hereafter, we call this energy shift energy. If directly interpreting it as a thermal energy at zero temperature, this result implies that there exists non-vanishing thermal energy even at zero temperature. This statement does not appear to make sense. Therefore, shift energy should be removed to obtain reasonable thermodynamic interpretation.
The shift energy can be eliminated by shifting the origin of thermodynamic energies, which is associated with taking an appropriate integral constant when integrating the first law of thermodynamics.
Another ambiguity occurs for $m =0$ (or $r_h = \sqrt{m_{\ph}}$). First, one can naively expect that  \eq{res:energieswithm} in the $m \to 0$ limit reproduces the results, \eq{res:interenm0} and  \eq{res:freeenm0}, obtained for $m=0$. However, it is not the case. In \eq{res:energieswithm} the internal energy becomes zero at $r_h \to \sqrt{m_{\ph}}$, while the internal energy studied in the previous section with $m=0$ does not vanish, $E = \fr{3 \pi^3 V}{2  G} T_H^4$. Why does this discrepancy occur in spite of the fact that $m$ and $r_h$ are continuous parameters? In subsequent sections, we will investigate how these unusual features can be resolved. 

Lastly, it is worth noting that there exists another way to understand physics of the dual QFT following the AdS/CFT correspondence. The holographic renormalization technique is a useful tool to extract much important information about the dual QFT. For a SAdS black hole with a planar boundary, it has been well known that the holographic renormalization leads to the exact same results obtained from the black hole thermodynamics. In more general cases, however, the holographic renormalization results do not usually coincident with those of black hole thermodynamics. This is also true for the above black hole composed of two matters. The internal energy derived from the holographic renormalization, as will be shown, is given by   
\be
E  = \fr{3 \ls 4 m + m_{\ph}^2 \rs}{64 \pi G } .
\ee
Naively comparing it with the black hole thermodynamic result in \eq{res:energieswithm},  one can easily see that the black hole thermodynamics and the holographic renormalization lead to different results. We will also clarify why this discrepancy occurs.

\section{AdS black holes with different boundary topologies}

Before resolving problems raised in the previous section, we first investigate AdS black hole with a nontrivial boundary topology. Intriguingly, an AdS black hole with a hyperbolic boundary topology, which is called a topological black hole, also shows similar problems although the origin of them are different. In this section, we first try to resolve problems of the topological black hole.

\subsection{AdS geometries with different boundary topologies}
The AdS$_{d+1}$ geometry, which is a negatively curved spacetime, is defined as a hypersurface in an one-dimensional higher flat spacetime denoted by ${\bf R}^{2,d}$. The proper distance in the ambient flat space is described by
\be
ds^2 = - dy_{-1}^2 - dy_0^2 + d y_1^2 + \cdots + dy_d^2 ,
\ee
which is invariant under the $SO(2,d)$ Lorentz symmetry. In the ambient space, the AdS$_{d+1}$ geometry appears as the hypersurface satisfying the following constraint
\be
- y_{-1}^2 - y_0^2 + y_1^2 + \cdots + y_d^2 = - R^2 ,
\ee
where $R$ indicates the AdS radius. Since this constraint does not break the Lorentz symmetry of the ambient spacetime, the resuting geometry also preserves the $SO(2,d)$ symmetry which is nothing but the isometry of the AdS$_{d+1}$ spacetime. Due to this reason, the AdS$_{d+1}$ geometry is called a maximally symmetric space.

In order to get the AdS$_{d+1}$ metric, we need to rewrite the metric of the ambient space in terms of coordinates of the AdS$_{d+1}$ spacetime. There exist various coordinatizations which allow different boundary topologies. One of the well-known AdS$_{d+1}$ metric is given as
\be			\la{res:AdSmetric}
ds^2 =   - \fr{r^2 f_k (r)}{R^2}   dt^2 +  r^2  d\S_{k}^2   +  \fr{R^2}{r^2 f_k (r)}  dr^2  ,
\ee
where $f_k (r)$ is given by
\be
f_k (r) =  1 + k  \fr{R^2}{r^2} ,
\ee
and $k$ is either $0$ or $\pm 1$ relying on the boundary topology. For $k=0$ $d\S_{k}^2$ represents the metric of a $(d-1)$-dimensional flat space ${\bf R}^{d-1}$, while it becomes the metric of a unit sphere ${\bf S}^{d-1}$ for $k=1$ or that of a hyperbolic space denoted by ${\bf H}^{d-1}$ for $k=-1$.

For more concreteness, let us further consider the explicit representation of AdS$_{d+1}$ \cite{Emparan:1999pm}. 
For $k=0$ one can use the following coordinatization  
\bea
y_{-1} + y_{d} = R^2 r \quad  {\rm and} \qquad y^\m = \fr{r}{R} x^\m \ (\m=0,1,\cdots,d-1)  .
\eea
From the constraint, one can easily find
\be
y_{-1} - y_d = \fr{1}{r} + \fr{r}{R^4} \et_{\m\n} x^\m x^\n ,
\ee
where $\et_{\m\n}$ denotes a $d$-dimensional Mikowski metric.
Substituting these relations into the metric of the ambient space, we finally obtain
\be
d s^2 =  - \fr{r^2}{R^2}  dt^2  +  r^2 \ls d u^2 + u^2 d\O_{d-2}^2 \rs + \fr{R^2}{r^2} dr^2 ,
\ee
where $u$ is dimensionless and the bulk spacetime is foliated with slices corresponding to the flat $d$-dimensional Minkowki spacetime. In the asymptotic region ($r \to \infty$), the boundary metric is given by a $d$-dimensional Minkowki metric up to the conformal factor, $r^2/R^2$. For $k=1$, the constraint can be satisfied by the following coordinatization
\bea
&& y_{-1} = \sqrt{r^2 + R^2} \cos\ls \fr{t}{R} \rs \, \quad y_0 = \sqrt{r^2 + R^2} \sin\ls \fr{t}{R} \rs \ , \nn
&&  y_1 = r \cos u \ , \quad y_2 =  r \sin u \cos \th_1 \ , \quad \cdots \nn
&& y_{d} = r \sin u \sin\th_1 \cdots \sin \th_{d-2} .
\eea
Using this, an induced metric on the hypersurface is reduced to
\be
ds^2 = - \fr{r^2}{R^2} f_1 (r) dt^2  +  r^2  d \O_{d-1}^2  + \fr{R^2}{r^2 f_1 (r)} dr^2 ,
\ee
where $d \O_{d-1}^2 = d u^2 + \sin^2 ud \O_{d-2}^2$ with $0 \le u< \pi$ indicates the metric of a $(d-1)$-dimensional sphere with a unit radius. Similarly, we can take the following parametrization for $k=-1$
\bea
&& y_{-1} = r \cosh u  \ , \quad y_0 = \sqrt{r^2 - R^2} \sinh \ls \fr{t}{R} \rs \ , \quad  y_{d} = \sqrt{r^2 - R^2} \cosh \ls \fr{t}{R} \rs \ , \nn
&&  y_1 = r \sinh u \cos \ph_1 \ , \quad y_2 =  r \sinh u \sin \ph_1 \cos \ph_2 \ , \quad \cdots \nn
&& y_{d-1} =  r \sinh u \sin \ph_1 \sin\ph_2 \cdots \sin \ph_{d-2} .
\eea
Then, the resulting AdS$_{d+1}$ metric becomes
\be
ds^2 = - \fr{r^2}{R^2} f_{-1} (r) dt^2 + r^2  dH_{d-1}^2 + \fr{R^2}{r^2 f_{-1} (r)} dr^2  ,
\ee
with
\be
dH_{d-1}^2 =du^2 + \sinh^2 u  \ d \O_{d-2}^2
\ee
where $ dH_{d-1}^2$ implies the metric of a $(d-1)$-dimensional hyperbolic space with a unit radius.
These AdS metrics with different topologies appear as a vacuum solution of a gravity theory with a negative cosmological constant
\be
S = \fr{1}{16 \pi G} \int d^{d+1} x  \sqrt{-g} \ \ls {\cal R} - 2 \L \rs ,
\ee
where
\be
\L = - \fr{d (d-1)}{2 R^2}  .
\ee

\subsection{Black hole thermodynamics with different boundary topolgies}

It is worth noting that there exist another vacuum solution when there is a matter localized at the center of the AdS space. This solution has been known as a black hole. The black hole metric has the same metric form in \eq{res:AdSmetric} with a black hole factor. If we denote the mass density of the localized matter as $m$, the general black hole geometry depending on $k$ can be classified by
\be
ds^2 =   - \fr{r^2 f_k (r)}{R^2}   dt^2 +   r^2  \ls d u^2 + \r_k^2 (u) \ d \O_{d-2}^2 \fr{}{}  \rs  +  \fr{R^2}{r^2 f_k (r)}  dr^2
\ee
with the following generalized metric factor
\be
f_k (r) =  1 + k \fr{R^2}{r^2} - \fr{m}{r^d} .
\ee
Here $\r_k (u)$ plays a role of radius of a $(d-2)$-dimensional sphere
\bea
\r_k &=& u  \quad  (  0 \le u < \infty ) \quad {\rm for} \ k=0   , \nn
	&=& \sin u \quad  (  0 \le u\le \pi )  \quad {\rm for} \ k=1, \nn
	&=&\sinh u \quad  (  0 \le u< \infty )  \quad {\rm for} \ k=-1.
\eea

It is well known that macroscopic quantities of a black hole can be reinterpreted as those of a thermal system. To see this, let us first define the event horizon where the black hole factor vanishes, $f(r_h) = 0$. Using this fact, the black hole mass density can be rewritten in terms of the black hole horizon
\be
m = r_h^d + k R^2 r_h^{d-2} .
\ee
Due to the absence of a conical singularity at the black hole horizon, the Hawking temperature is determined to be
\be
T_H = \fr{1}{4 \pi R^2} \ls d r_h + \fr{(d-2) k R^2}{r_h} \rs .
\ee
The area law of the Bekenstein-Hawking entropy leads to the following thermal entropy
\be
S_{BH} = \fr{ r_h^{d-1} V_k}{4 G}  ,
\ee
where $V_k$ indicates the area independent of the radial coordinate. Following the AdS/CFT correspondence, $V_k$ can be reinterpreted as the spatial volume of the dual field theory. It usually diverges except the case with $k=1$, so we need to regularize it. For $k=0$ and $k=-1$, we restrict the region of $u$ as $0 \le u \le u_0$. Then, $V_k$ is given by
\bea
V_0 &=&  \fr{2 \pi^{(d-1)/2}}{\G\ls \fr{d-1}{2} \rs} \ \int_0^{u_0} \d u \ u^{d-2} \quad {\rm for} \ k=0 , \nn
V_1 &=& \fr{2 \pi^{(d-1)/2}}{\G\ls \fr{d-1}{2} \rs} \ \int_0^{\pi} du \ \ls \sin  u \rs^{d-2} =  \fr{2 \pi^{d/2}}{\G\ls d/2\rs}   \quad {\rm for} \ k=1 , \nn
V_{-1} &=& \fr{2 \pi^{(d-1)/2}}{\G\ls \fr{d-1}{2} \rs} \ \int_0^{u_0} du \ \ls \sinh  u \rs^{d-2}  \quad {\rm for} \ k=-1  .
\eea

Integrating the first law of thermodynamics, the internal energy reads
\be  \la{res:thermoinenergy}
E_{bh}  = \fr{(d-1) V_k}{16 \pi G R^{2}} \ls r_h^d + k R^2 r_h^{d-2} \rs  = \fr{(d-1)   m V_k}{16 \pi G R^{2}}  ,
\ee
and the free energy, $F = E -T_H S_{BH}$, is given by
\be   \la{res:thermofreeenergy}
F_{bh} = - \fr{ V_k}{16 \pi G R^{2}} \ls r_h^d -  k R^2 r_h^{d-2} \rs   .
\ee
Note that, although these quantities satisfy the first law of thermodynamics, there still exists an ambiguity related to an integral constant. To resolve such an ambiguity, let us concentrate on the zero temperature limit of the above quantities.
For $k=0$ and $k=1$, the zero temperature limit appears at $m=0$ (or $r_h=0$). When the above black hole's energies are identified with those of a thermal system, they automatically vanish at zero temperature. This fact implies that an additional integral constant is not required to interpret the black hole quantities as thermal quantities. This is not true for $k=-1$ due to the nontrivial topology. In this case, the geometry still remains as a black hole even for $m=0$, which is called a topological black hole. In addition, there is no $r_h=0$ limit.

In order to understand thermodynamics of the topological black hole correctly, we need to study the black hole geometry in more detail. For simplicity, we focus on the $d=4$ case where the black hole factor reduces to
\be
f_{-1} = 1 - \fr{R^2}{r^2} - \fr{m}{r^4} .
\ee
This metric shows that even for $m=0$ there exists a black hole horizon at $r_h = R$. For $m>0$,
there exists one real root similar to the previous black hole with a non-negative $k$. Even for $- R^4/4 \le m<0$ unlike the previous cases, there exists a black hole solution which has two horizons, an inner and outer horizon, like a charged black hole. Especially, $m=-R^4/4$ describes the extremal limit corresponding to zero temperature. The black hole thermodynamics leads to the following internal and free energy 
for the topological black hole 
\bea			\la{res:nkmnotzero}
E_{bh}  &=& \fr{3V_{-1}}{16 \pi G R^{2}} \ls r_h^4 -  R^2 r_h^{2} \rs   , \nn
F_{bh} &=& - \fr{ V_{-1}}{16 \pi G R^{2}} \ls r_h^4 + R^2 r_h^{2} \rs  .
\eea
Before understanding these black hole quantities, it should be noted that they can not be applied to
the case with $m =0$. To see this, let us first consider the case with $m=0$.
At the horizon ($r_h=R$), the Hawking temperature and Bekenstein-Hawking entropy are given by
\be
T_H = \fr{r_h}{2 \pi R^2}  \quad {\rm and} \qquad S_{BH} = \fr{V_{-1}}{4 G} r_h^3 .
\ee
Using the first law of thermodynamics, the internal and free energy read
\bea		\la{res:m0energies}
E^0_{bh}  &=& \fr{3V_{-1}  }{32 \pi G R^{2}}  r_h^4   , \nn
F^0_{bh} &=& - \fr{ V_{-1}}{32\pi G R^{2}}  r_h^4  .
\eea
Comparing these results with those of \eq{res:nkmnotzero} in the $r_h \to R$ limit, one can easily see that the black hole thermodynamics for $m=0$ is not continuously connected to the $m \ne 0$ case (see Fig. 1). In order to understand thermodynamics of the topological black hole correctly, we need to figure out why this discrepancy happens. In addition, there exists another obstacle which makes the thermodynamic interpretation vague even for $m \ne 0$. In the zero temperature limit ($r_h = R/\sqrt{2}$), the black hole's internal and free energy do not vanish
\bea		
F_{bh} = E_{bh}  = - \fr{3 V_{-1} R^{2}}{64 \pi G }  .
\eea
If  black hole's energies are naively identified with thermal energies, this result implies that there exist non-zero thermal energies even at zero temperature. This is happen because the origin of the thermal energies is shifted due to the nontrivial topology. For obtaining a well-defined thermodynamics, we need to define a shift energy as
\be
E_{sh}  \equiv  - \fr{ 3 V_{-1} R^{2}}{64 \pi G } ,
\ee
and to remove it from the black hole energies. This shift energy can be removed by taking a proper integral constant when we perform the integration of the first thermodynamic law. Then, thermal energies for $m \ne 0$, which automatically vanish at zero temperature, can be defined as
\bea
E_{th} &=& E_{bh} - E_{sh}  = \fr{3V_{-1}}{64 \pi G R^{2}} \ls 4 r_h^4 -  4 R^2 r_h^{2} +  R^4 \rs   , \nn
F_{th} &\equiv& E_{th} - T_H S_{BH}=  F_{bh} - E_{sh} =  - \fr{ V_{-1}}{64 \pi G R^{2}} \ls 4 r_h^4 + 4 R^2 r_h^{2} - 3  R^4\rs  .
\eea

\begin{figure}
\begin{center}
\vspace{0cm}
\hspace{-0.5cm}
\subfigure[internal energy]{ \includegraphics[angle=0,width=0.45\textwidth]{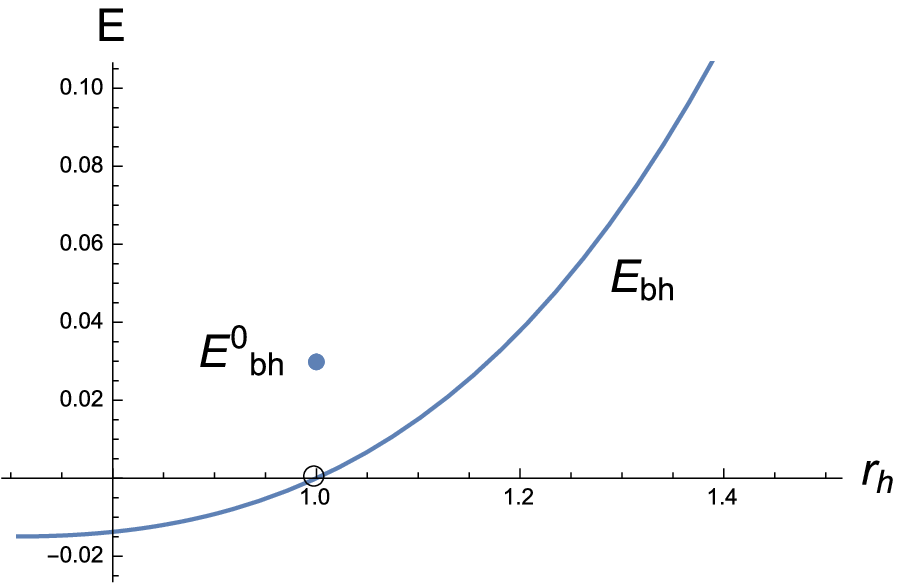}}
\hspace{0cm}
\subfigure[free energy]{ \includegraphics[angle=0,width=0.45\textwidth]{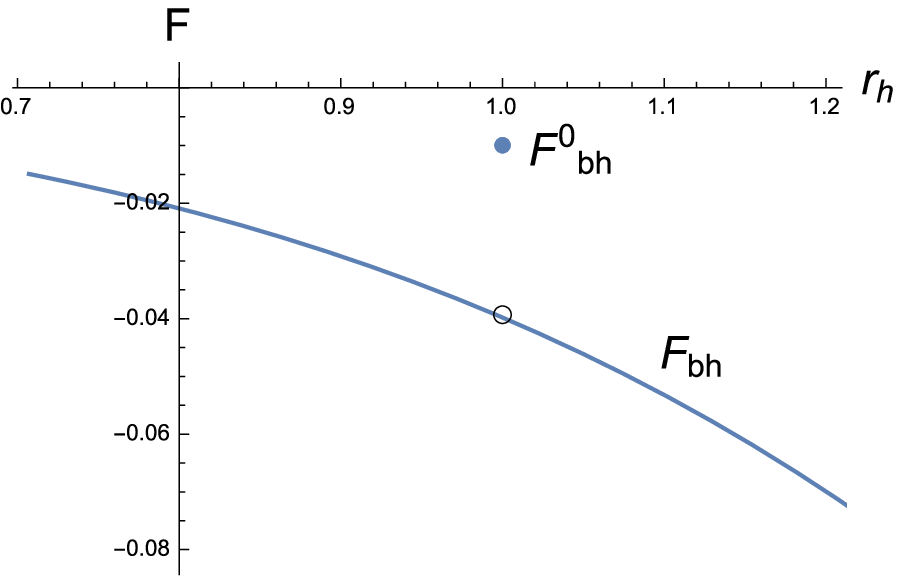}}
\vspace{-0cm}
\caption{\small  The internal and free energies for $m \ne 0$ (solid lines) and $m = 0$ (represned by dots) where we set $R=1$. Above the empty dot on the energy curves implies that energies are not defined on that point.}
\label{number}
\end{center}
\end{figure}

\subsection{ Holographic renormalization for AdS black hole}

Following the AdS/CFT correspondence, black hole thermodynamics can be reinterpreted as that of the dual field theory defined at the AdS boundary. In the AdS/CFT context, the well-known method to obtain physical quantities of the dual field theory is to apply the holographic renormalization technique \cite{de Boer:1999xf,deHaro:2000vlm,Skenderis:2002wp,Balasubramanian:1999re}. In this procedure, we need counterterms to remove the divergences of the on-shell gravity action. Depending on the dimension of the AdS$_{d+1}$ spacetime, the required counterterms are different \cite{Balasubramanian:1999re,Emparan:1999pm}. In this work, we will focus on the $d=4$ case. It is straightforward to generalize to other dimensions.
In order to clarify the finite temperature field theory dual of the AdS black hole, we use the Euclidean signature with which the gravity action becomes
\be
S_G = - \fr{1}{16 \pi G} \int_{ {\cal M}} d^{5} x  \sqrt{g} \ \ls {\cal R} - 2 \L \rs ,
\ee
where the metric is given by
\be			\la{res:EAdSmetric}
ds^2 =  \fr{R^2}{r^2 f_k (r)}  dr^2 + \fr{r^2}{R^2} \ls f_k (r) dt^2 + d\S_{k}^2 \rs  .
\ee
In general, the variation of the gravity action is not well defined at the boundary, so that the well-defined variation requires the Gibbons-Hawking term at the boundary \cite{Gibbons:1976ue}
\be
S_{GB} =  \fr{1}{8 \pi G} \int_{\pa {\cal M}} d^{4} x \sqrt{\g} \ {\cal K} ,
\ee
where $\g$ is an induced metric on the boundary and ${\cal K}$ indicates the trace of the extrinsic curvature, ${\cal K}_{\m\n} = - \fr{1}{2} \ls  \nabla_\m n_\n + \nabla_\n n_\m\rs$, with a unit normal vector $n_{\m}$. According to the AdS/CFT correspondence, the on-shell gravity action is identified with the generating functional of the dual quantum field theory. Since the on-shell gravity action usually diverges at the asymptotic boundary, we need additional boundary terms called the counterterm, which get rid of the divergences of the on-shell gravity action \cite{Balasubramanian:1999re,Emparan:1999pm}
\be
S_{ct} = \fr{1}{8 \pi G} \int_{\pa {\cal M}}  d^4 x \sqrt{\g} \ls \fr{3}{R} + \fr{R}{4} {\cal R}^{(4)} \rs ,
\ee
where ${\cal R}^{(4)}$ means the intrinsic curvature scalar of the four-dimensional boundary. As a consequence, the renormalized action $S_{re}$ is given by
\be		\la{res:freeenergy}
S_{re} \equiv \b F = S_G + S_{GH} + S_{ct} ,
\ee
where $\b$ and $F$ imply the inverse temperature and free energy, respectively. From this finite free energy, one can easily derive the boundary stress tensor \cite{Balasubramanian:1999re,Emparan:1999pm,Park:2014gja,Park:2015afa}
\bea		\la{res:boundstressten}
T_{\m\n} \equiv - \fr{2}{\sqrt{\g}} \fr{\d F}{\d \g^{\m\n}}= \fr{1}{8 \pi G} \lb - {\cal K}_{\m\n} + \g_{\m\n} {\cal K} + \fr{3}{R} \g_{\m\n} - \fr{R}{2} \ls {\cal R}^{(4)}_{\m\n} - \half \g_{\m\n} {\cal R}^{(4)} \rs \rb .
\eea
Then, the internal energy and pressure in an $i$-direction are defined as
\bea
E_{HR} &=& \int d^3 x \sqrt{\g} \g^{00} T_{00} , \la{res:intenergy} \\
P^i_{HR} &=& - \int d^3 x \sqrt{\g} \g^{ii} T_{ii}  ,
\eea
where $i$ in the last relation is not summed.

\subsubsection{For $k=0$}

Rewriting the black hole mass in terms of the horizon, the black hole metric can be rewritten as
\be
f_0 (r) = 1 - \fr{r_h^4}{r^4} .
\ee
Substituting this metic into formulas in \eq{res:freeenergy} and \eq{res:intenergy}, we obtain the free and internal energy in terms of the horizon
\bea
F_{HR} &=& - \fr{u_0^3  }{12  G  R^2} \ r_h^4  , \nn
E_{HR} &=& \fr{3   u_0^3}{12 G R^2} \ r_h^4 .
\eea
These results exactly coincide with those obtained from black hole thermodynamics, as mentioned before.
The pressure of this system is given by
\be
P^i_{HR} = \fr{  u_0^3 }{12 G R^2 } \ r_h^4  .
\ee
Note that the free and internal energy as well as the pressure vanishes in the zero temperature limit ($r_h \to 0$) and that the trace of the stress tensor also vanishes even at finite temperature.

\subsubsection{For $k=1$}

Now, repeat the previous calculation for $k=1$. The holographic renormalization yields the following quantities
\bea
F_{HR} &=& - \frac{\pi }{8 G R^2}\ls  r_h^4  -  R^2 r_h^2 \rs  +\frac{3 \pi  R^2}{32 G} , \nn
E_{HR} &=& \frac{3 \pi }{8 G R^2} \ls  r_h^4 + R^2 r_h^2 \rs +\frac{3 \pi  R^2}{32 G} , \nn
P_{HR}^i &=& \frac{\pi}{8 G R^2} \ls  r_h^4 + R^2 r_h^2 \rs +\frac{ \pi  R^2}{32 G} .
\eea
Comparing these results with the previous thermodynamic results in \eq{res:thermoinenergy} and \eq{res:thermofreeenergy}, one can see that they give different results unlike the previous planar case with $k=0$.  Why does this discrepancy appear?
In general, the holographic renormalization maps the calculations of a classical gravity to nonpertubative ones of the dual quantum field theory. This means that the holographic renormalization results include all quantum and thermal corrections.
To extract the quantum effect from the holographic renormalization result, one can simply take the zero temperature limit.
Unlike the planar case, the zero temperature quantities for $k=1$ do not vanish, which can be interpreted as the vacuum quantities. As a consequence, we can divide the holographic renormalization result into vacuum and the black hole contributions 
\bea
F_{HR} &=& F_{vac} + F_{bh} , \nn
E_{HR} &=& E_{vac}  + E_{bh} .
\eea
For $k=0$ the vacuum contributions, $E_{vac}$ and $F_{vac}$, vanish because of the trivial boundary topology. For $k=1$, however, the nontrivial boundary topology gives rise to nontrivial vacuum contribution  corresponding to the Casimir effect \cite{Balasubramanian:1999re}
\be
E_{vac} = F_{vac} = \frac{3 \pi  R^2}{32 G} .
\ee
Subtracting the vacuum contribution from the holographic renormalization results, the resulting energies, $F_{bh}$ and $E_{bh}$, become perfectly coincident with the results obtained by the black hole thermodynamics in \eq{res:thermoinenergy} and \eq{res:thermofreeenergy}. 
In sum, $E_{vac}$ and $F_{vac}$ vanish for $k=0$, while they for $k=1$ have a non-vanishing value corresponding to the Casimir effect. Removing this vacuum contribution from the holographic renormalization results, the remaining energies, $F_{bh}$ and $E_{bh}$, are consistent with the black hole thermodynamics  and can be directly reinterpreted as thermal energies of the dual QFT.

\subsubsection{For $k=-1$}

For $k=-1$, the dual Euclidean quantum field theory is defined on ${\bf S} \times {\bf H}^{3}$. Unlike the previous cases, the range of the radial coordinate of the black hole geometry is restricted to $r \ge r_h $ even for $m \to 0$.
Using this lower bound, the holographic renormalization gives rise to the following free and internal energy
\bea
F_{HR} &=& \frac{\left(4 m+3 R^4-8 r_h^4\right) (\sinh (2 u_0)-2 u_0) }{64 G R^2} , \nn
E_{HR} &=& \frac{3 \left(4 m+R^4\right) (\sinh (2 u_0) - 2 u_0 )}{64 G R^2} , \nn
P_{HR}^i &=& \frac{ \left(4 m+R^4\right) (\sinh (2 u_0) - 2 u_0 )}{64 G R^2} ,
\eea
which are not the same as the free and internal energy obtained from the black hole thermodynamics.
In order to understand the holographic renormalization result, let us first recall that the holographic result can be represented as $F_{HR}= F_{vac} + F_{bh}$ and $E_{HR}= E_{vac} + E_{bh}$, as shown in the previous section. Noting that the vacuum energies are given by the part independent of the black hole horizon
\be
E_{vac} = F_{vac} = \frac{ 3 R^4 \lb \sinh (2 u_0)-2 u_0 \rb  }{64 G R^2} .
\ee
we can easily see that $F_{bh}$ and $E_{bh}$ perfectly match to the thermodynamic result in \eq{res:nkmnotzero}. Remembering that $F_{bh}$ and $E_{bh}$ for $k=-1$ can be further decomposed into $F_{bh}=F_{th} + F_{sh}$ and $E_{bh}=E_{th} + E_{sh}$ due to the shift caused by the hyperbolic boundary topology, energies obtained by the holographic renormalization are finally rewritten as
\bea
F_{HR} &=& F_{th} + F_{vac} + F_{sh} , \nn
E_{HR} &=& E_{th}+ E_{vac} + E_{sh} .
\eea
where thermal energies, $F_{th}$ and $E_{th}$, should vanish at zero temperature. This fact requires that the vacuum energies at zero temperature must cancel the shift energy
\be
E_{vac} = F_{vac} = -   E_{sh} = \fr{ 3 V_{-1} R^{2}}{64 \pi G } .
\ee
Due to this relation, the holographic energies can be identified with the thermal energies directly. These results explain what the difference between the holographic renormalization and black hole thermodynamics is. The holographic renormalization gives rise to more general result involving the vacuum contribution. Ignoring such a vacuum contribution, the holographic renormalization leads to the same result obtained by the black hole thermodynamics.

In the previous section, we showed that topological black hole's thermodynamics with a nonzero $m$
does not approach to the one with $m=0$ in the $m \to 0$ limit. In order to understand why this discordance happens, one should first notice that the AdS radius $R$ for $m=0$ becomes the horizon, so that it is not a constant independent of thermodynamics for $m=0$. This means that the vacuum energies for $m \ne 0$ become thermal energies at $m = 0$. In other words, there is no vacuum energy, $E_{vac} = F_{vac}=0$. As a consequence, the resulting thermal energies for $m=0$ should be given by
\bea
F^0_{th} &=&  F_{HR}  -  F_{sh}  = -  \frac{ R^4 (\sinh (2 u_0)-2 u_0) }{32 G R^2} , \nn
E^0_{th} &=& E_{HR} - E_{sh}  =  \frac{3 R^4  (\sinh (2 u_0) - 2 u_0 )}{32 G R^2} ,
\eea
where $V_{-1} = \pi \ls \sinh(2 u_0) -2 u_0 \rs$. This result perfectly coincides with the black hole's thermodynamic energies obtained in \eq{res:m0energies} with $m=0$.

\section{Black hole composed of two matters}

Now, return to the black hole composed of two matters. In Sec. 2, some issues related to it have been raised. In Sec. 3, we showed that the similar problems also occurs for the AdS black hole with a hyperbolic boundary topology and clarified why such a discrepancy happens. In this section, we will resolve issues raised in Sec. 2 by applying the same method used in the topological black hole.

Let us first investigate the holographic renormalization of the previous Einstein-scalar gravity which yields the energies involving the vacuum as well as thermal contributions. To do so, we consider the Euclidean version of \eq{act:Einsteinscalar}
\be
S_G = - \fr{1}{16 \pi G} \int d^{5} x \sqrt{g} \lb {\cal R} - 2 \L - \half \d_{ab} \
\pa_\m \ph^{a} \pa^\m \ph^{b} \rb
\ee 
In order to get a well-defined variation and the finite on-shell gravity action, we need the Gibbons-Hawking term 
\be
S_{GB} =  \fr{1}{8 \pi G} \int_{\pa {\cal M}} d^{4} x \sqrt{\g} \ {\cal K} ,
\ee
together with the following counterterms \cite{Fukuma:2002sb,Nojiri:1999jj} 
\be
S_{ct} = \fr{1}{8 \pi G} \int_{\pa {\cal M}}  d^4 x \sqrt{\g} \ls \fr{3}{R}  - \fr{1}{8}   \d_{ab} \ \pa_\m \ph^a  \pa^\m  \ph^b \rs ,
\ee
where the last term is required to get rid of the additional UV divergence caused by the scalar field.
Then, the renormalized action, $S_{re} \equiv S_G + S_{GB} + S_{ct}$, leads to the following free energy
\be
F_{HR} \equiv \fr{S_{re}}{\b} = \frac{ \ls -8 r_h^4+3 m_{\phi }^2+4 m \rs V}{64 \pi  G},
\ee 
where $V$ indicates a regularized spatial volume of the dual QFT. From it, the stress tensor of the dual field theory reads \cite{Balasubramanian:1999re,Emparan:1999pm,Park:2014gja,Park:2015afa}
\bea		
T_{\m\n} \equiv - \fr{2}{\sqrt{\g}} \fr{\d F}{\d \g^{\m\n}}= \fr{1}{8 \pi G} \lb - {\cal K}_{\m\n} + \g_{\m\n} {\cal K} + \fr{3}{R} \g_{\m\n} + \fr{1}{4} \d_{ab} \pa_\m \ph^a \pa_\n \ph^b - \fr{1}{8} \g_{\m\n}  \d_{ab} \pa_\r \ph^a \pa^\r \ph^b \rb .
\eea
The internal energy and pressure of the dual field theory are given by
\bea
E_{HR} &\equiv& \int d^3 x \sqrt{\g} \g^{00} T_{00} = \fr{3\ls 4 m + m_{\ph}^2 \rs V }{64 \pi G }, \nn
P^i_{HR} &\equiv& - \int d^3 x \sqrt{\g} \g^{ii} T_{ii}  = \fr{ \ls 4 m + m_{\ph}^2 \rs V}{64 \pi G }, 
\eea
Similar to the topological black hole for $k=-1$, the holographic renormalization result shows a different result from the the black hole thermodynamics in \eq{res:energieswithm}. In order to understand it, we first need to decompose it to the vacuum and black hole contribution. The vacuum contribution is independent of the black hole horizon, so that it reads
\be
F_{vac} = E_{vac}  =  \frac{   3 m_{\phi }^2    V}{64 \pi  G}, 
\ee
which has the same magnitude as the shift energies defined in \eq{res:shiftenergy} with an opposite sign, $F_{vac} = E_{vac}=-F_{sh} = -E_{sh}$. Subtracting the vacuum energies from the holographic energies, we arrive at
\bea
F_{bh} &=& F_{HR} - F_{vac} = - \frac{  V}{16 \pi  G} \ r_h^2 \ls r_h^2 + m_{\ph}\rs , \nn
E_{bh} &=& E_{HR} - E_{vac} = \fr{3  V }{16 \pi G }  \ r_h^2 \ls r_h^2 - m_{\ph}\rs,
\eea
where $m=r_h^2 \ls r_h^2 - m_{\ph}\rs$ is used. This result perfectly matches to the black hole's thermodynamic result in \eq{res:energieswithm}.
Similar to the topological black hole, the above black hole has also a nontrivial energies even at zero temperature. In order to define thermal energies properly, we should remove them by shifting the origin of the thermal energies. Then, thermal energies can be defined as
\bea 		\la{res:thermalenergiesmne0}
F_{th} &=& F_{HR} - F_{vac} - F_{sh} =F_{bh} - F_{sh} ,\nn
E_{bh} &=& E_{HR} - E_{vac} - E_{sh} =E_{bh} -E_{sh} .
\eea 
These thermal energies automatically vanish at zero temperature.

Now, let us move to the case with $m=0$. In the $m \to 0$ limit, the above result in \eq{res:thermalenergiesmne0} does not approach to the results in \eq{res:interenm0} and \eq{res:freeenm0}, as mentioned before. This is because in the $m \to 0$ limit the black hole horizon is related to $m_{\ph}$. Therefore, the vacuum energies described by $m_{\ph}$ becomes thermal energies. As a consequence, the resulting thermal energies at $m \to 0$ reduce to
\bea			\la{res:freeenm01}
F_{th} &=&  F_{HR} - F_{sh} = - \fr{ V}{32 \pi G} \ m_{\ph}^2,\nn
E_{bh} &=&E_{HR}  - E_{sh} = \fr{3 V}{32 \pi G}  \ m_{\ph}^2 ,
\eea
which are again in agreement with the thermodynamic results with $m=0$, \eq{res:interenm0} and \eq{res:freeenm0}.

As shown above, there exist two different back holes with the momentum relaxation. If there is no localized matter ($m=0$), the thermal free energy of the black hole only with the momentum relaxation  leads to
\be
F_{th} = - \fr{ \pi^3 V}{2  G} \ T_H^4 ,
\ee
where the temperature is fixed by $T_H = \sqrt{m_{\ph}}/2 \pi$. For the black hole composed two kinds of matter $(m \ne 0)$, its thermal free energy reduces to
\be			\la{res:freeenergymne0}
F_{th} =  -\frac{V T_H \lb \pi ^2 T_H^2 \left(\sqrt{\pi ^2 T_H^2+2 m_{\phi }}+\pi  T_H\right)+m_{\phi } \left(2 \sqrt{\pi ^2 T_H^2+2  m_{\phi }}+3 \pi  T_H\right)\rb}{32 G} .
\ee
Now, let us compare these free energies at a fixed temperature. Since the temperature is fixed by $m_{\ph}$ for $m=0$, the fixed temperature implies that $m_{\ph}$ is also fixed to be $m_{\ph} = 4 \pi^2 T_H^2$. Substituting these relations into the free energy into \eq{res:freeenergymne0}, it reduces to 
\be
F_{th} = - \fr{ 40 \pi^3 V}{32  G} \ T_H^4 .
\ee
Since the free energy for $m \ne 0$ is  always smaller than that for $m=0$, the black hole composed two kinds of matter is preferred at a given temperature.


\section{Discussion}

We have studied thermodynamics of a scalar hairy black hole, which is dual to a strongly interacting QFT with a momentum relaxation. Intriguingly, this scalar hairy black hole showed exotic thermodynamic features. Its thermodynamics defined with $m \ne 0$ does not smoothly approach to the one defined with $m=0$ even though all parameters are continuous. In addition, its thermal energies do not vanish even at zero temperature. These atypical features occur because of the existence of the secondary hair. In general, a primary hair of a black hole is associated with a certain conserved quantity, so it usually generates a physically well-defined thermal quantity. However, it is not clear what the corresponding conserved charge is for a secondary hair. In several different scalar hairy black holes, it has been shown that a secondary hair just modifies a black hole mass and the power of the radial coordinate dependence. Despite such changes, its thermodynamic law is not altered because the secondary hair does not generate an additional thermodynamic variable. Due to this reason, we have reinvestigated thermodynamics including the scalar hair related to a momentum relaxation. After calculating the boundary stress tensor by using the holographic renormalization, which describes quantum and thermal effects at the same time, we properly defined correct thermal energies by subtracting the vacuum contribution. Intriguingly, the resulting energies are perfectly matched to those of black hole's thermodynamics obtained by assuming the first law of thermodynamics.
 
We also discovered the reason why black hole thermodynamics with $m \ne 0$ is not continuously connected to the case with $m=0$. For $m \ne 0$, the vacuum energy is independent of the black horizon, so that thermal energy can be well defined by subtracting the vacuum energy from the holographic renormalization result and gives rise to the consistent thermodynamic result. For $m = 0$, however, it is not the case because the vacuum energy is directly connected with the black hole horizon. Therefore, it must be understood as the thermal energy rather than the vacuum energy. This prescription accounts for why thermodynamic interpretation jumps at a specific parametric line with $m=0$. In the topological AdS black hole, the similar features also happen. We showed that the same method can explain the similar exotic features of the topological black hole. The resulting thermodynamic interpretation is perfectly coincident with the black hole thermodynamics derived from the first law of thermodynamics. If one meets a black hole solution having a similar problem, the study of this work would be helpful to understand its black hole thermodynamics. Finally, we have compared the free energy of two black holes derived from the same Einstein-scalar gravity theory. At a given temperature, we showed that the black hole with $m\ne0$ is always preferable to one with $m=0$.

\vspace{1cm}

{\bf Acknowledgement}

C. Park was supported by Basic Science Research Program through the National Research Foundation of Korea funded by the Ministry of Education (NRF-2013R1A1A2A10057490) and also by the Korea Ministry of Education, Science and Technology, Gyeongsangbuk-Do and Pohang City.

\vspace{1cm}


\end{document}